\documentclass{llncs}
\usepackage{graphicx}
\usepackage[cmex10]{amsmath}
\usepackage{xspace}
\usepackage{amssymb}
\usepackage{tikz}
\begin{document}
\title{Knowledge is at the Edge! How to Search in Distributed Machine Learning Models}
\titlerunning{Confidence-Driven Knowledge Retrieval}  

\author{Thomas Bach \and Muhammad Adnan Tariq \and Ruben Mayer \and Kurt Rothermel}

\authorrunning{Bach et al.} 

\institute{Institute of Parallel and Distributed Systems, University of Stuttgart, Germany,\\
\email{$\langle$firstname.lastname$\rangle$@ipvs.uni-stuttgart.de},\\ \texttt{http://www.uni-stuttgart.de}}

\maketitle              

\begin{tikzpicture}
\begin{scope}[overlay]
\footnotesize
\node[text width=40cm] at ([yshift=-17.0cm,xshift=8cm]current page.south) {\copyright 2017 Springer International Publishing AG. This is the authors' version of the work. \\The definite version is published in OTM 2017 Conferences / CoopIS 2007.  \\The final publication is available at Springer via https://doi.org/10.1007/978-3-319-69462-7\_27};
\end{scope}
\end{tikzpicture}

\begin{abstract}

With the advent of the internet of things and industry 4.0 an enormous amount of data is produced at the edge of the network. Due to a lack of computing power, this data is currently send to the cloud where centralized machine learning models are trained to derive higher level knowledge. With the recent development of specialized machine learning hardware for mobile devices, a new era of distributed learning is about to begin that raises a new research question: How can we search in distributed machine learning models? Machine learning at the edge of the network has many benefits, such as low-latency inference and increased privacy. Such distributed machine learning models can also learn personalized for a human user, a specific context, or application scenario. 
As training data stays on the devices, control over possibly sensitive data is preserved as it is not shared with a third party. This new form of distributed learning leads to the partitioning of knowledge between many devices which makes access difficult. In this paper we tackle the problem of finding specific knowledge by forwarding a search request (query) to a device that can answer it best. To that end, we use a entropy based quality metric that takes the context of a query and the learning quality of a device into account.
We show that our forwarding strategy can achieve over 95\,\% accuracy in a urban mobility scenario where we use data from 30\,000 people commuting in the city of Trento, Italy.

\end{abstract}


\newcommand{\Entity}{\ensuremath{N}\xspace}
\newcommand{\Context}{\ensuremath{\omega}\xspace}
\newcommand{\Query}{\ensuremath{\overrightarrow{q}}\xspace}
\newcommand{\RoutingTable}{\ensuremath{RM}\xspace}
\newcommand{\Hops}{\ensuremath{H}\xspace}
\newcommand{\Result}{\ensuremath{R}\xspace}
\newcommand{\Quality}{\ensuremath{Q}\xspace}
\newcommand{\History}{\ensuremath{\overrightarrow{\Entity}}\xspace}
\newcommand{\Similarity}{\ensuremath{S}\xspace}

\section{Introduction}\label{sec:intro}

In many areas such as stock trading, drug design, manufacturing, and urban mobility \cite{domingos:2012,kienzle:2016} machine learning is the key enabler of optimization and driver of performance \cite{domingos:2012,manyika:2011,khan:2012}. Besides choosing the right machine learning algorithm and applying it right, the amount of training data is key to success \cite{domingos:2012}. While the selection and application of machine learning algorithms is a research field of its own, enough training data is needed to calibrate machine learning models, such that they can make correct predictions. 

With the advent of paradigms like the Internet of Things, smart city, and Industry 4.0, data will be abundantly available \cite{manyika:2011}. Cisco, for example, estimates that the I.o.T. alone will generate over 400 ZB of data annually, by 2020 \cite{Cisco:2015}. In particular, the proliferation of smart phones made training data from different sensors, such as accelerometers, cameras, microphones, and GPS units widely available \cite{khan:2012}. Google reported that by centralizing a great amount of training data for speech recognition from Google voice search \cite{schalkwyk:2010}, it became possible to train high-quality feature-rich machine learning models for voice recognition\cite{heigold:2013}.

The current approach to share such information is massive centralization. In many application scenarios, however, centralization of possibly sensitive data is not desirable as centralized data is regularly subject to breaches \cite{informationisbeautiful:2015}. Today, it is well known that it is possible to derive knowledge of a user's habits, such as his home and work location from his GPS traces \cite{Ashbrook:2003}. Many human users are thus unwilling, at least uncomfortable sharing such private information \cite{Ziegeldorf:2014}. 

The common approach to tackle this issue is to distribute the computing infrastructure \cite{mayer:2015}, and even push computing towards the edge of the network \cite{garcia:2015,montresor:2016}. The upcoming trend of \emph{fog computing}\cite{hong:2013,Mayer:2017} supports this by providing computational resources close to the edge, creating a computational continuum that spans from the edge devices to the centralized cloud data centers. Sensitive data can then be processed directly on devices that are under control of the user or on fog nodes very close to them. In this respect, mobile device manufactures are building specialized machine learning hardware that enables machine learning at the edge\footnote{Mark Gurman; BloombergTechnology, Apple Is Working on a Dedicated Chip to Power AI on Devices: https://www.bloomberg.com/news/articles/2017-05-26/apple-said-to-plan-dedicated-chip-to-power-ai-on-devices}. Machine learning at the edge has many additional advantages, it allows for example to keep the user in the loop, learn personalized,  and offer low latency feedback \cite{garcia:2015,montresor:2016}. Google for example has recognized this trend and made approaches, where personalized learning is done directly on smart phones \cite{McMahan:2017}; however, the generated local machine learning models are synchronized with a central server. Google argues, that by processing the data locally, privacy is increased compared to an entirely centralized approach. 

Completely decentralized learning also holds great challenges. As training data is not centralized, each machine learning model is only trained with respect to its local experiences. In particular, such models may become local experts that are very good in predicting local phenomena. In a medical scenario this might be an advantage, as a model could learn the peculiarities of one specific patient and enable a detailed analysis. For other use cases, this is not enough. In an urban mobility scenario, for example, users are usually more interested, in traffic conditions in another part of the city which they have never seen before. Distributed learning holds the opportunity to learn about local phenomena in great detail on the one hand, on the other hand it creates the problem of locating specific knowledge.

To address this problem, in this paper, we present methods to route a query for specific knowledge through a network of nodes (local experts) that each train a local machine learning model. Our goal is to forward such a query to the node that can answer it best. In particular, we look at scenarios where knowledge in the form of machine learning models is fully distributed. Such a fully decentralized approach holds three mayor difficulties: First, we cannot assume a central index of all available knowledge. Second, the different devices (nodes) might learn based on different local observations and contexts. Third, parts of the knowledge changes or becomes outdated over time.




To this end, our contributions are: (1) We propose a decentralized routing strategy that forwards queries for specific knowledge towards nodes that can answer them best. (2) We propose methods to maintain routing tables that guide the forwarding of such a query. (3) We use entropy to evaluate how good a given query can be answered based on its context and the local machine learning model of a node. (4) We develop a modified form of the Barabasi Albert model \cite{albert:2002} to generate a scale free topology that clusters network nodes with similar knowledge close together to deal with heterogeneous knowledge. With its scale free properties such a topology provides short paths between any two network nodes and is robust against node failures. (5) We show that we can achieve over 95\,\% accuracy when using synthetic data and data generated by a mobility simulator where 30\,000 people commute in the city of Trento, Italy, in the context of different weather conditions, times of the day, and traffic conditions.




\section{System Model and Problem Formulation}\label{sec:SystemModel}

We assume a distributed system of fog \cite{hong:2013,Mayer:2017} nodes that each train a machine learning model based on local observations. These nodes can join and leave the system at any time and range from user managed devices such as smart phones, laptops, and desktop computers, to infrastructure based services located in data centers, such as private clouds. All nodes communicate directly over a undirected, scale free topology, i.e. power law distributed node degree and short paths. These properties make scale free networks particularly well suited for our problem as they connect two arbitrary nodes (e.g. the source and optimal destination of a query) with a small number of hops. Furthermore, many existing networks such as social networks or the internet already show scale free properties \cite{amaral:2000}. Maintaining such a topology is also a well studied research problem \cite{Li:2004,albert:2002}. 

\begin{figure}
\includegraphics[width=\linewidth]{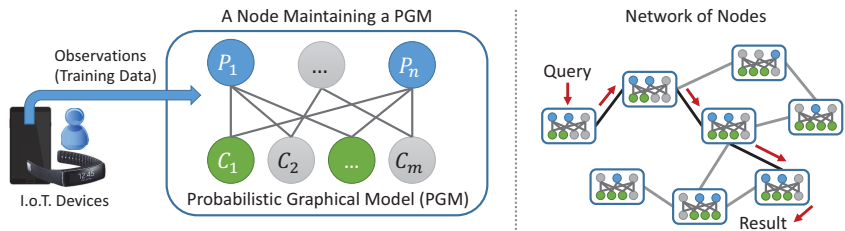}
\caption{System overview.}
\label{fig:SMOverview}
\end{figure}

In order to learn, all nodes maintain a graphical machine learning model as shown in Fig.~\ref{fig:SMOverview}. Graphical models (Probabilistic Graphical Models, PGM) such as Bayesian networks or conditional random fields have a wide range of machine learning applications in computer vision, natural language processing, and bioinformatics \cite{sutton:2006}. In a PGM, random variables are represented as nodes and dependencies between them as edges of a graph. This gives them great flexibility in modeling complex dependencies.

We assume, that all network nodes maintain structural identical PGMs that are continuously evolving based on individual training data (observations). This training data can be generated either by the nodes themselves, e.g. a smart phone generates GPS traces from its internal sensors, or can be received from other sensors, such as a wristband sensor that collects cardiovascular data. Furthermore, the different nodes learn about different phenomena in different contexts, leading to individual expertise of the different nodes. In an urban mobility scenario, for example, two nodes could learn about traffic conditions in different parts of the city at different times of the day. 
In particular, this means that the different nodes train different subsets of random variables. In consequence, not all nodes can predict all random variables equally well. A reliable prediction about the outcome of a specific random variable thus requires to search for (i.e. query) the network node that has best training.

Given any PGM, we categorize the random variables of the PGM into two groups: predicting variables and context variables. Predicting variables are the subset of random variables that we want to predict based on a certain context, modeled as context variables. In an urban mobility scenario, where we want to predict the travel time for the streets in a city with a PGM, the travel time for each street would be represented as a predicting variable. Factors that influence this travel time, such as weather or time of the day would be represented as context variables. 
In Fig.~\ref{fig:SMOverview} this categorization is reflected by the color of the random variable. Predicting variables $P_n$ in blue, context variables $C_n$ in green and untrained random variables of both types in grey. 

In this context, we define a query as a request to predict the outcome for a specific predicting variable in a certain context. In this respect, we define context as given assignments for a set of context variables. Our goal is to forward a query to a node that can answer it with the highest possible quality (we introduce quality in Sec:~\ref{sec:confidenceMetric}).

\subsection{Formal Model and Problem Statement}
More formally, we assume a set of network nodes $\Entity = \{\Entity_1, ... , \Entity_n\}$ that are connected over a scale free topology. Each node $\Entity_x$ holds a PGM that consists of a graph $G=(V,E)$ of discrete random variables $V$ where dependencies between random variables are modeled by the edges $E$. We classify the random variables into predicting variables $P_n = \{p^{1}_{n}, ..., p^{m}_{n}\}$ and context variables $C_{n} = \{c^{1}_{n}, ..., c^{m'}_{n}\}$, where $p_n$ and $c_{n}$ are possible assignments. Each random variable must be classified either as predicting variable or as context variable ($V = C \cup P$ and $C \cap P = \emptyset$).

We define a query $\Query = (P_x,\{c_{x}^{y}, ... ,c_{x'}^{y'}\},\Hops,\Result,\Quality,\History)$,
where $P_x$ is the random variable that needs to be predicted in context of a given set of assignments ($\{c_{x}^{y}, ... ,c_{x'}^{y'}\}$) for a subset of context variables and a limited number of hops \Hops (number of times a query can be forwarded). Furthermore, the query contains a field to hold the prediction result \Result, the estimated quality of this result \Quality and a vector of visited nodes \History. 


We can now define the concrete knowledge retrieval problem. Given i) a set of nodes $\{\Entity_1, ... , \Entity_n\}$ holding ii) continuously evolving, heterogeneously trained PGMs and iii) a query \Query for specific knowledge, our goal is to maximize the retrieval quality of a query while forwarding it only \Hops times.

In the following, we first establish a notion of knowledge quality in the context of PGMs and describe how to measure the quality with that a node can answer a query. (cf. Sec.~\ref{sec:confidenceMetric}). Based on this quality metric, we present methods to route a query towards the node that can answer it with highest quality in Sec.~\ref{sec:routing}.

\section{Entropy, a Measure of Training Quality}\label{sec:confidenceMetric}
In this section we discuss how we measure the training quality of a PGM. Based on this quality, we describe how to estimate the quality with that a PGM can answer a query for specific knowledge. 

As stated above, Probabilistic Graphical Models (PGM) consist of interdependent random variables. Such models are usually designed by an expert who puts his domain knowledge in the structure of the model, e.g. chooses the random variables and their conditional dependencies such that the model reflects the dependencies in the real world. Training data is then used to converge the probability distributions of the random variables from a uniform distribution to the distributions of the real world. In other words, if an increasing amount of training data is fed into the machine learning model, the uncertainty of the model decreases. In machine learning, this uncertainty (or often called surprise) of a model is measured by calculating the entropy of its random variables \cite{manning:1999}. 

\begin{figure}
\includegraphics[width=\linewidth]{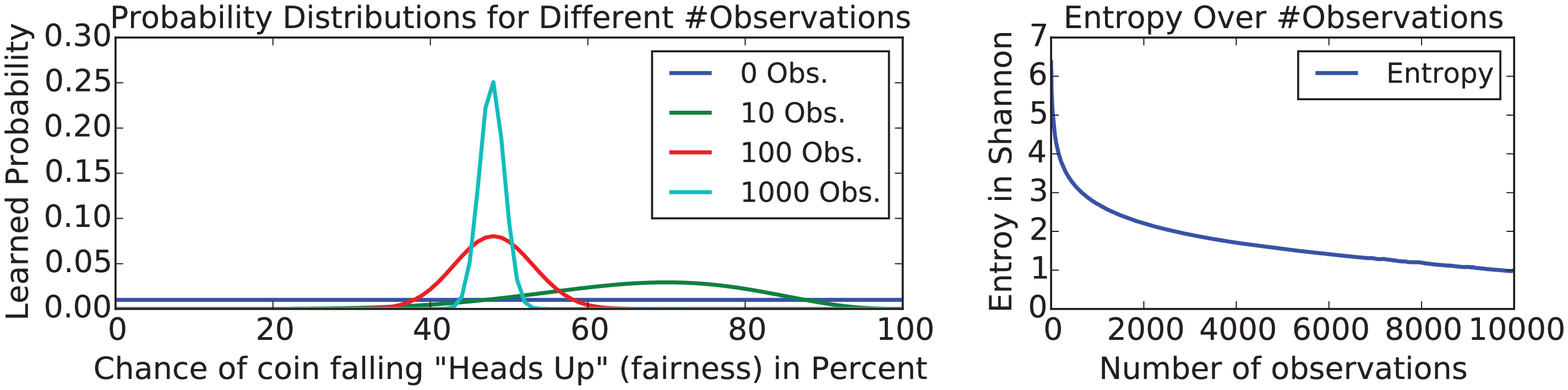}
\caption{Probability distribution (l) and entropy (r) of a coin flip for different number of observations.}
\label{fig:entropy}
\end{figure}

If, for example, we want to learn the probability of a coin flip being ``Heads Up'', we could use a very simplistic model that only consists of one random variable $X$ with possible outcomes $\{0\%, ..., 100\%\}$. We now flip a fair coin several times and use the results to train the random variable as shown in Fig.~\ref{fig:entropy}. With an increasing number of observations (or coin tosses) the ``true'' probability distribution establishes and the entropy decreases.

Given a random variable $X$ with possible assignments $\{x_1, ..., x_n\}$ we can calculate the entropy ($H(X)$) as the average surprise (or uncertainty) of the random variable (cf. Eq\,\ref{eq:entropy}). The logarithm of the probability of an assignment $log(P(x_n))$ represents the amount of surprise we perceive for the specific outcome \cite{manning:1999}. The ``surprises'' of all possible outcomes are then weighted by their probability $P(x_n)$ and summed up to one entropy value often also called self-information \cite{manning:1999}. 

\begin{equation}
\label{eq:entropy}
H(X) = - \sum_{k=1}^{n} P(x_k)log_2 P(x_k)
\end{equation} 

In complex machine learning models, we usually want to predict the outcome of multiple random variables. In these cases we calculate the joint entropy $H(X_0,...,X_n)$ in order to describe their ``joint uncertainty'', e.g. $\{X_0,...,X_n\}$ (cf. Eq.\,\ref{eq:jointEntropy}). Similar to entropy for one random variable, the idea is to calculate the uncertainty for each combination of random variables involved and weight these combinations w.r.t. their probabilities.

\begin{equation}
\label{eq:jointEntropy}
H(X_0,...,X_n) = - \sum_{x_0\in X_0}\cdot\cdot\cdot\sum_{x_n\in X_n} P(x_0,...,x_n) log_2 P(x_0,...,x_n)
\end{equation} 

The joint entropy describes the ``total uncertainty'' of multiple random variables. Given a random variable $X_0$ that is dependent on the outcome of other random variables $\{X_1,...,X_n\}$, the entropy $H(X_0,...,X_n)$ denotes the uncertainty of the outcome given that we don't know anything about the outcome of $\{X_0,...,X_n\}$. If we now gain information about the outcome of one of the variables (e.g. $X_0$ is a context variable and its outcome is given by a query), we can derive the remaining entropy (uncertainty) according to the chain rule of conditional entropy by subtracting the entropy $H(X_0)$ from the total entropy $H(X_0,...,X_n)$, cf. Eq.\,\ref{eq:condEntropy}. In the following, we use this chain rule to calculate the uncertainty which the PGM of a specific network node has to answer a query.

\begin{equation}
\label{eq:condEntropy}
H(X_1,...,X_n|X_0) = H(X_0,...,X_n)-H(X_0)
\end{equation} 

In Sec.~\ref{sec:SystemModel} we divided the random variables of a PGM in two categories, predicting variables and context variables, where each predicting variable is dependent on the outcomes of a number of independent context variables. For a given predicting variable $P_0$ that is dependent on the outcome of context variables $\{C_0,C_1,C_2\}$ we can calculate the joint entropy $H(P_0,C_0,C_1,C_2)$ and individual entropies for the context variables $H(C_0),H(C_1),H(C_2)$. Given a query $\Query = (P_0,\{c_{0}^{1},c_{1}^{3}\},...)\footnote{For better readability we do not state all fields of the query here (i.e. \Hops, \Result, \Quality, \History).}$ for $P_0$ with observed outcomes $c_{0}^{1}\in C_0$ and $c_{1}^{3}\in C_1$ we can calculate the remaining uncertainty of the PGM to answer the query by subtracting the entropy of the context variables from the joint entropy of the predicting variable (cf. Eq.\,\ref{eq:entropyQuery}). This results in the remaining uncertainty of the PGM to answer the query.

\begin{equation}
\label{eq:entropyQuery}
H(P_0,C_3|C_0,C_1) = H(P_0,C_0,C_1,C_2) - H(C_0) -H(C_1)
\end{equation}

For the rest of this paper we will also refer to entropy as the learning or training quality of a PGM.

\section{Routing}\label{sec:routing}

Now that we have established how we can measure the training quality of the PGMs of each node, we describe how we build routing models and use them to forward queries towards the node that can answer them best. 
In contrast to a classic routing table, where a network address is associated with a specific port (outgoing link), each node $\Entity_i$ maintains a routing model $\RoutingTable_{\Entity_n}^{\Entity_i}$ for each neighbor $\Entity_n$. Each $\RoutingTable_{\Entity_n}^{\Entity_i}$ serves as a descriptive model that cumulatively represents the knowledge available over the respective outgoing link. Keeping link-individual routing models is necessary, because we need to calculate the estimated answering quality of a query with respect to its context (cf. Sec.\,\ref{sec:confidenceMetric}) as storing all possible combinations of contexts easily becomes too much overhead.

In order to process a query, a node first tries to improve the prediction \Result of a query based on its local PGM. In the next step, the node uses its routing models to determine to which neighbor the query should be forwarded. As this is an approximate routing process, we limit the number of hops (\Hops) that a query \Query is forwarded before the result is returned to the sender. 

In order to improve the retrieval quality we use a network topology that clusters nodes with similar knowledge (nodes that have learned about a similar set of predicting nodes). We can then optimize our routing models by maintaining context information of predicting variables only for predicting variables that have been learned by the cluster. This leads to a double-staged routing approach, where a query for a predicting variable $P_n$ is first forwarded to a cluster of nodes that have learned about $P_n$. In the second stage, the query is then forwarded within a cluster to a node that has learned it in the requested context. In the following we describe how we build the routing tables, forward a query, maintain the network topology and deal with loops in the topology in detail.

\subsection{Building the Routing Tables}
The routing models $\RoutingTable_{\Entity_n}^{\Entity_i}$ that each node $\Entity_i$ maintains for every neighbor $\Entity_n$ store entropy values of random variables (predicting variables and context variables) and represent the knowledge available to the respective neighbor $\Entity_n$. 
In order to maintain them in a proactive fashion, each node sends summaries of its entropy values stored in its local PGM and its routing models to its neighbors, whenever they have changed above a certain threshold and a minimum amount of time has passed since the last update. This makes sure that all neighbors have up-to-date information about their neighbors and at the same time avoids that the network is flooded with updates. 

\begin{figure}
\includegraphics[width=\linewidth]{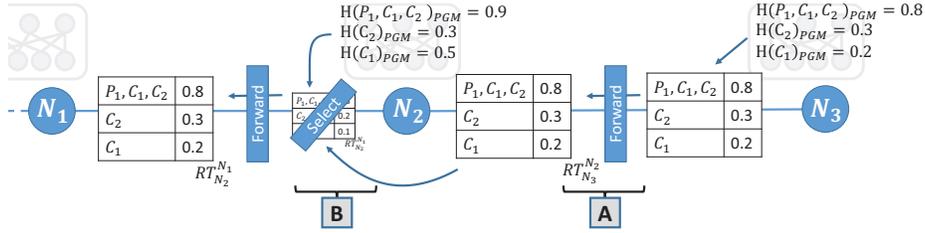}
\caption{Forwarding entropy values in same context.}
\label{fig:routingSimple}
\end{figure}

In the following, we explain this forwarding process with respect to a set of nodes $\{\Entity_1,\Entity_2,\Entity_3\}$ in detail. For better readability and without loss of generality this example is with respect to one predicting variable $P_0$ and context variables $\{C_1, C_2\}$. A further simplification is the use of a flat topology, i.e. all nodes are connected in a line (cf. Fig.\,\ref{fig:routingSimple}). In the given example, $\Entity_3$ has only one neighbor ($\Entity_2$) and therefore can directly forward its set of entropy values $\{H(P_0, C_1, C_2)_{PGM},$ $ H(C_1)_{PGM},$ $H(C_2)_{PGM}\}$ from its PGM to its neighbor $\Entity_2$, where they are used as entries in the routing table $\RoutingTable_{\Entity_3}^{\Entity_2}$ (cf. Fig.\,\ref{fig:routingSimple} A). These entropy values represent the learning quality for $P_0$ available over the edge $\Entity_2\rightarrow\Entity_3$. As described in Sec.\,\ref{sec:confidenceMetric} this values can be used to calculate the quality with which a query in any context (i.e. {$\{\{C_0\},\{C_1\},\{C_0,C_1\}\}$} can be answered. 

\vspace{-1em}
When $\Entity_2$ has received the set of entropy values from $\Entity_3$, it decides to update the entropy values send to $\Entity_1$. In contrast to $\Entity_3$, $\Entity_2$ cannot send the entropy values from its PGM directly as it has to consider the entropy received from $\Entity_3$. Node $\Entity_2$ needs to select which set of entropy values ($\{H(P_0, C_1,$ $ C_2),$ $H(C_1),$ $H(C_2)\}$) is forwarded. In order to determine this, it compares the joint entropy value of its local PGM ($H(P_0, C_1, C_2)_{PGM} = 0.9$) with the joint entropy value of its routing table $\RoutingTable_{\Entity_2}^{\Entity_1}$ ($H(P_0, C_1, C_2)=0.8$) As the local joint entropy is higher ($0.9>0.8$) it forwards the complete entropy set for $P_0$ received from $\Entity_3$ ($\{H(P_0, C_1, C_2)=0.8, H(C_1)=0.2,H(C_2)=0.3\}$) to $\Entity_1$ (cf. Fig.\,\ref{fig:routingSimple} B). 

\textbf{Generalization:} If, in contrast to our example, a node has multiple neighbors, it stores the entropy value sets it receives from them in a separate routing table for each neighbor. In order to decide which entropy set (e.g. $\{H(P_0, C_1, C_2),$ $H(C_1),$ $H(C_2)\}$) should be forwarded, we compare all the entropy sets, including the entropy of the local PGM by their joint entropy values (i.e. $H(P_0, C_1, C_2)$) and forward the set with the lowest joint entropy. This forwarding approach makes sure that for each predicting variable ($P$), the lowest joint entropy (e.g. $H(P_0, C_1, C_2) = 0.8$) and the entropy values of its corresponding context variables (e.g $H(C_1)= 0.2,H(C_2)=0.3$) are propagated.

\begin{figure}
\includegraphics[width=\linewidth]{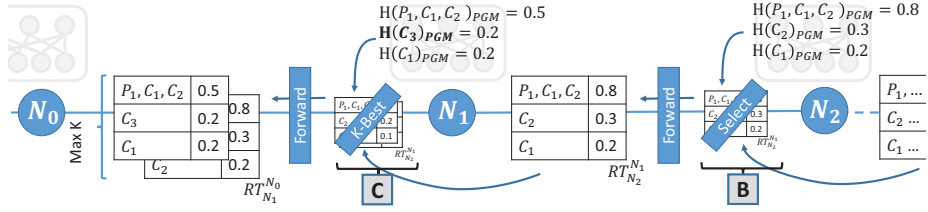}
\caption{Forwarding entropy values in different context.}
\label{fig:routingKBest}
\end{figure}

In cases where predicting variables have been trained with respect to different context variables by different nodes, entropy values cannot simply be merged. For example, $\Entity_2$ has trained $P_1$ with respect to context variables $\{C_1,C_2\}$ and $\Entity_1$ has trained $P_1$ with respect to $\{C_1,C_3\}$ (cf. Fig.\,\ref{fig:routingKBest}). In such cases, we store and forward up to $K \in \mathbb{N}$ different context combinations for each predicting variable $P$ (cf. Fig.\,\ref{fig:routingKBest} C). In this respect, $K$ is a design parameter that determines how many context combinations for one predicting variable are stored in the routing tables. In general $K$ is dependent on the number of relevant contexts in a concrete scenario. In cases where we have to limit them we can use existing dimension reduction algorithms to select the most important context variables. We will discuss the influence of $K$ in our evaluations in Sec.\,\ref{sec:evaluation}.

\subsection{Forwarding of a Query}
As discussed in Sec.\,\ref{sec:SystemModel}, a query \Query is a message issued by one node in the network, to retrieve a prediction for a specific predicting variable $P_x$ in a given context, represented by a set of assignments ($\{c_{x}^{y}, ... ,c_{x'}^{y'}\}$) for a subset of context variables $\{C_x,...,C_{x'}\} \in C$. A query is forwarded from one node to another until the predefined number of hops, \Hops, has been reached. 

When a node $\Entity_i$ receives a query \Query it first decreases the hop counter \Hops of the query and then determines if the query can be improved by the local PGM, by computing the entropy of answering the query (cf.\,Sec.\,\ref{sec:confidenceMetric}). The resulting entropy value is then compared to the entropy value \Quality in the query. If the entropy value in the query is higher than the locally computed value (i.e. the node has less uncertainty cf. Sec.\,\ref{sec:confidenceMetric}), the node predicts the outcome of the query with its PGM and updates the result field \Result and the quality field \Quality of the query \Query accordingly. If the hop counter \Hops of the query is greater then zero ($\Hops > 0$) the node uses its local routing models to select a neighbor to which the query is forwarded. 

In order to select a neighbor to send the query to, node $\Entity_i$ compares all routing models $\RoutingTable_{\Entity_n}^{\Entity_i}$ by computing the conditional entropy $H(P_x|C_x,...,C_{x'})$. This is done by subtracting the entropy values of the context variables ($\{H(C_x),$ $...,$ $H(C_{x'})\})$ from the joint entropy value $H(P_x, C_x, ..., C_{x'})$ stored in the routing tables $\RoutingTable_{\Entity_n}^{\Entity_i}$ (cf.\,Sec.\,\ref{sec:confidenceMetric}). The query is then forwarded to the neighbor with the smallest conditional entropy $H(P_x|C_x,...,C_{x'})$. 

\textbf{Discussion:} So far, we have described how routing models are built and how they are used to forward a query. The maintenance of entropy values in multiple routing models, especially for different context combinations, produces significant overhead. The number of possible context combination grows according to the binomial coefficient. If, for example, the nodes learns w.r.t. $5$ out of $10$ possible context variables, there are already $252$ possible combinations. In order to reduce this overhead, we cluster nodes that have learned about a similar set of predicting variables. Based on this clusters, our routing protocol uses two optimizations. First, nodes only forward entropy values from their PGM if they have a minimum level of quality (i.e. the entropy value is below a certain threshold). Second, if a node has not reached a minimum quality for a predicting variable $P_n$ it only maintains a single joint entropy value for $P_n$ (no entropy values for context) in its routing models. This single joint entropy value can then be used to forward a query to the next cluster that has learned about $P_n$, where context sensitive routing, as described above, is performed. In the following we describe how we maintain such a clustered network topology.




\subsection{Topology Maintainance}\label{sec:Topology}
As mentioned in previous sections, the topology of our network should exhibit scale free properties, such as a power law distributed node degree and short paths. Additionally we want to cluster nodes that have learned about a similar subset of predicting variables. In order to manage the overhead of maintaining routing models for each neighbor, we also need to  give each node the option to limit the maximum number of neighbors. This limit can be determined node-individually, e.g., dependent on the amount of memory consumed by the routing models. In order to generate such a topology, we use a modified version of the Barabasi Albert model \cite{albert:2002}. Our idea is to make the preferential attachment of the Barabasi Albert model dependent on node similarity. The original algorithm starts with an initial set of $m_0$ nodes and connects a new node $\Entity_n$ to an existing node $\Entity_e$ with a probability proportional to the edge degree of the existing nodes. This way, $\Entity_n$ can connect with up to $m<m_0$ existing nodes. In the original algorithm the probability of a node $\Entity_x$ connecting to an existing node $\Entity_y$ is given by 
$p_{x\rightarrow y}=\frac{k_y}{\sum_j k_j}$ where $k_y$ is the degree of the existing node divided by sum of all edge degrees. We multiply this probability with a similarity factor $\Similarity(PGM_{\Entity_x},PGM_{\Entity_y})\rightarrow[0,1]$ that describes the similarity between two PGM (e.g. $PGM_{\Entity_x}$ and $PGM_{\Entity_y}$). If a node already has reached its individual maximum number of edges ($edgelimit$) we set the probability to zero as shown in Eq.\,\ref{eq:preferentialProbability}.

\begin{equation}\label{eq:preferentialProbability}
p_{x\rightarrow y}= 
\begin{cases}
\frac{k_y}{\sum_j k_j} \cdot{•} Similarity(PGM_{\Entity_x},PGM_{\Entity_y}) &\quad \text{if } k_y \leq edgelimit\\
0 &\quad \text {else} \\
\end{cases}
\end{equation}


\begin{figure}
\includegraphics[width=\linewidth]{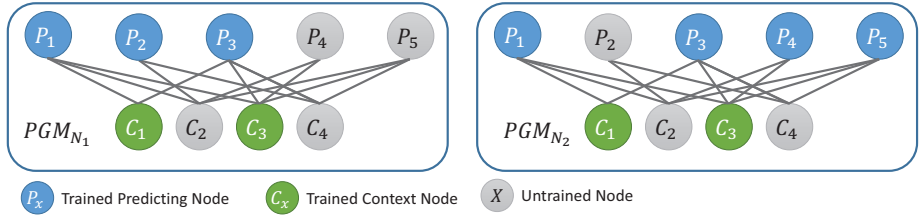}
\caption{Two PGMs with similar learning.}
\label{fig:Similarity}
\end{figure}

Let there be two nodes $\{\Entity_1,\Entity_2\}$ where $\Entity_1$ has trained the set $A=\{P_1, P_2, P_3\}$ and $\Entity_2$ the set $B=\{P_1, P_3, P_4, P_5\}$ of predicting variables of their PGM as shown in Fig.\,\ref{fig:Similarity}. We define the similarity between them as the size of the intersection between $A$ and $B$ divided by the minimum cardinality of $A$ and $B$ cf.\,Eq.\,\ref{eq:similarity}.

\begin{equation}
\Similarity(PGM_{\Entity_1},PGM_{\Entity_2}) = \frac{|A \cap B|}{ \min(|A|,|B|)}= \frac{2}{ \min(3,4)} = \frac{2}{3}
\label{eq:similarity}
\end{equation}


To demonstrate that this modified algorithm still produces a topology with power law distributed node degree, we plotted the number of edges for a network of 600 nodes using the original Barabasi Albert Model and our modified version where we limited the number of edges to 60. The major difference is, that our modified version exhibits multiple nodes degree 60 instead of having several nodes with degree $>$ 60.

\begin{figure}
\includegraphics[width=\linewidth]{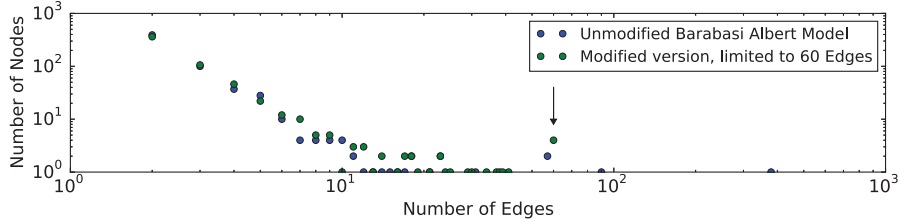}
\caption{Comparison of node degree between the modified and unmodified B. A. Model.}
\label{fig:NodeDegree}
\end{figure}
In order to deal with loops in the topology we slightly increase the forwarded entropy value with every hop. This way, propagated entropy values increase more over longer propagation paths than over shorter ones. Queries will then always be forwarded over shortest paths with lower entropy values. 

\section{Evaluation}\label{sec:evaluation}
In this section we evaluate the above presented aggregation based routing strategy (ABS) with respect to different network sizes, number of context nodes, context combinations, number of hops, and the context diversity parameter K (cf. Sec.\,\ref{sec:routing}) on synthetic training data and on data from an urban mobility scenario, in the following referred to as ``Trento data''. We compare our strategy to a directed random walk approach commonly used in unstructured Peer to Peer networks cf.\,Sec.\,\ref{sec:RelatedWork}.

We implemented our routing strategy (cf.\,Sec.\,\ref{sec:routing}) in the Peer to Peer simulator PeerSim \cite{Montresor:2009} and performed our evaluations on an Open Stack virtual machine with 64 cores and 256\,GB RAM running Ubuntu 16.04. We used PeerSim to instantiate up to $32\,768$ ($2^{15}$) network nodes. To represent the Probabilistic Graphical Models (PGM), each node individually trained a Bayesian network consisting of an experiment dependent number of random variables. In the following we will state for each experiment how many predicting variables ($P_n$) and context variables ($C_n$) were used to form a PGM as described in Sec.\,\ref{sec:SystemModel}. The Bayesian networks on the individual nodes were then trained  either on synthetic training data (Gaussian distributed observations) or on data from the Trento data set respectively. The use of synthetic data gave us the ability to flexibly generate experiment setups with any number of predicting or context variables. Based on the trained nodes, the topology between the nodes was created as described in Sec.\,\ref{sec:routing}. 

Based on this setup, we used the cycle based engine of PeerSim to perform our evaluations. In each cycle, each node first propagated its knowledge and then issues a query that is forwarded as described in Sec.\,\ref{sec:routing}. In order to determine the accuracy of the result, we compared the entropy of the result with the optimal entropy which was determined by an exhaustive search over all nodes.

The ``Trento data'' data originates from a real world simulator for collaborative and distributed learning \cite{poxrucker:2014} developed at the German center for artificial intelligence (DFKI) to generate large-scale, realistic data sets for machine learning. The simulator is based on the city map of Trento in Italy. It features genuine bus tables, weather, and commuting statistics of the city. Based on this data, we used the simulator to hosts 30\,000 autonomous agents that emulate the behavior of citizens, even forming traffic jams that lead to different travel times at different times of the day, days of the week under different weather conditions for different road segments of the Trento street graph. In our experiments we used weather, time of the day, and day of the week as context nodes to predict travel times for different road segments that we used as predicting variables.

\begin{figure}[h!]
\centering
\includegraphics[width=\linewidth]{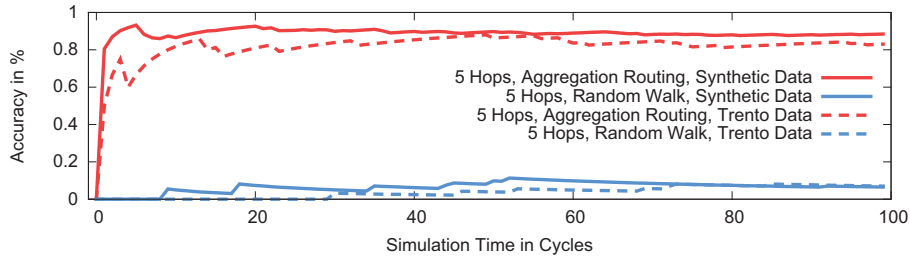}
\caption{Retrieval accuracy averaged over time (cycles) compared.}
\label{fig:SynthReal}
\end{figure}

In our first evaluation (Fig.\,\ref{fig:SynthReal}) we compare the retrieval accuracy of our aggregation based strategy (ABS) on synthetic data and the Trento data averaged over an increasing number of cycles with the random walk approach. In this evaluation we used $1024$ nodes that we trained on 100 different predicting variables and 3 context variables (weather, time of the day, and day of the week in the Trento data). We can see, that in the first cycle, the accuracy is low, as most of the routing models are empty. As knowledge gets propagated through the network, the retrieval quality increases until it settles around 90\%. This evaluation already indicates the good performance of our algorithm on synthetic data and on the Trento data.

\begin{figure}[h!]
\centering
\includegraphics[width=1\linewidth]{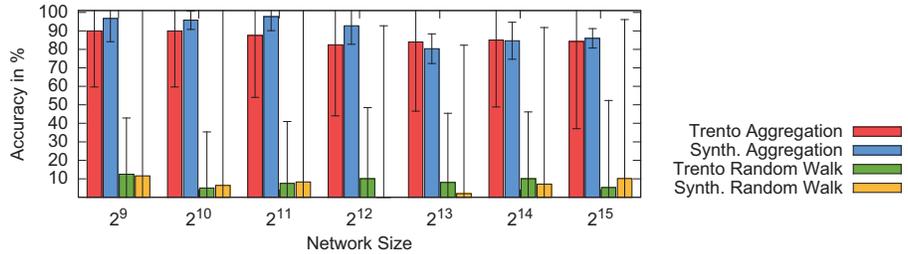}
\caption{Network size and retrieval quality.}
\label{fig:scalability}
\end{figure}

In Fig.\,\ref{fig:scalability} we evaluate the performance of ABS at different network sizes (up to $32\,768$ nodes) with the random walk approach using synthetic and Trento data. Just like in the previous evaluation we used 100 predicting nodes and 3 context nodes according to the Trento data. We forwarded each query $2 \cdot log(network\,size)$ times. In comparison to the synthetic data, the standard deviation (indicated by the whiskers in the graph) is a bit higher for the Trento data. The reason for this is, that the knowledge about some predicting variables is scarce and thus harder to find.

\vspace{-3em}

\begin{figure}[h!]
\centering
\includegraphics[width=1\linewidth]{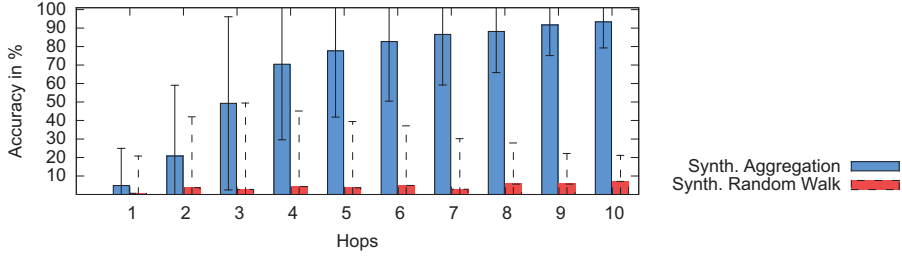}
\caption{Retrieval Quality w.r.t different number of hops.}
\label{fig:Hops}
\end{figure}

Fig.\,\ref{fig:Hops} shows the influence of the number of hops. For this evaluation we used a network of 512 nodes, 10 predicting variables, and 3 context variables on synthetic data. We can see that with an increasing number of hops not only the accuracy increases, but also the standard deviation (whiskers) decreases. In general the number of required hops grows proportional to the network diameter. As we are using a free scale topology, this is approximately logarithmic to the number of nodes (cf.\,Sec.\,\ref{sec:routing}).
 
\begin{figure}[h!]
\centering
\includegraphics[width=\linewidth]{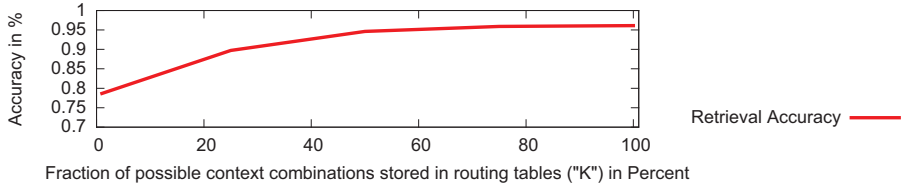}
\caption{Retrieval quality with respect to $K$.}
\label{fig:KValue}
\end{figure}

In the following, we will have a closer look at the influence of the number of possible combinations of context variables and their influence on routing accuracy. We introduced the problem of different context combinations in Sec.\,\ref{sec:routing} and tackled it by introducing a parameter $K$ that defines how many different context combinations are stored in the routing models. In Fig.\,\ref{fig:KValue} we can see that keeping about 50\% of all possible context combinations already leads to a reasonably good retrieval accuracy. 

\begin{figure}[h!]
\centering
\includegraphics[width=1\linewidth]{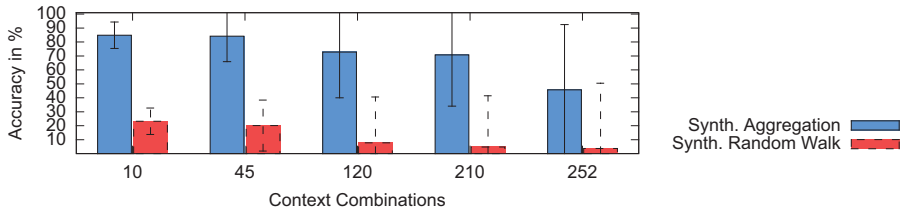}
\caption{Possible context combination for $K=10$, 512 nodes, and 3 hops.}
\label{fig:ContextCombinations}
\end{figure}

Fig.\,\ref{fig:ContextCombinations} shows how the accuracy degrades with an increasing amount of context combinations for a network of $512$ nodes, $K=10$, one predicting variable, and $10$ context variables of which up to 4 have been trained. According to the binomial coefficient this creates up to $252$ possible context combinations that could have been learned. We can see that with an increasing amount of possible context, not only the accuracy decreases, but also the standard deviation of the accuracy increases. The surprisingly good performance and low standard deviation for the random strategy for $10$ context combinations can be explained when realizing that there are potentially up to 51 nodes that have learned the respective contexts. 
\vspace{-1em}

\vspace{-3em} 
\begin{figure}[h!]
\centering
\includegraphics[width=1\linewidth]{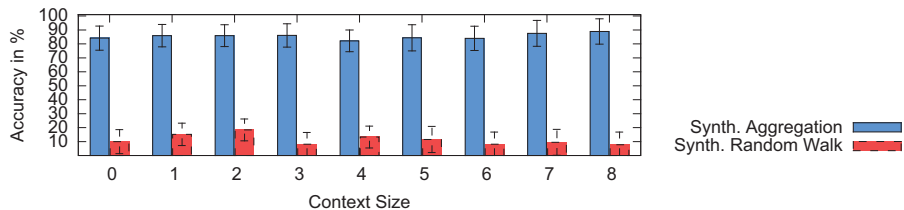}
\caption{Retrieval accuracy with respect to different number of context nodes}
\label{fig:ContextSize}
\end{figure}

\vspace{-1em}
When $K$ is chosen around 50\% of the number of context combinations, retrieval accuracy is fairly independent form the number of context variables used, as shown in Fig.\,\ref{fig:ContextSize}. For this experiment we used a network of $2048$ nodes that were trained on 500 predicting variables in up to $10$ contexts, $25$ context combinations and $k=12$. As we highlighted in Sec.\,\ref{sec:routing} we can use of the shelf methods for dimension reduction to determine most important context combinations w.r.t. the application scenario at hand.

\section{Related Work}\label{sec:RelatedWork}

Information retrieval from peer to peer (P2P) systems and machine learning are both well studied areas. Today, machine learning is often done in Big Data scenarios, where all training data is logically centralized. There exist many approaches to distribute the training data and machine learning models between several machines, for example on a cluster. These systems have the benefit of a centralized controller that actively manages how information is distributed between the different machines. In such scenarios, communication-efficient distribution of data between machines is a hard research problem of its own \cite{Mayer:2016}.

In this paper we argue that with the trend of decentralized computing, machine learning is coming to the edge of the network (cf.\,Sec.\,\ref{sec:intro}). On the one hand, this enables many benefits such as low latency access and the ability to maintain control over sensible information. On the other hand, without a central index structure, the problem of searching in distribute machine learning models is created. Therefore we focus on related work that tackles the problem of content sharing in P2P networks and discuss how fit these approaches are for knowledge retrieval. 



First P2P systems such as Chord \cite{Stoica:2001}, CAN \cite{Ratnasamy:2001} and Pastry \cite{Rowstron:2001} tackled the problem of how to find specific data items in a distributed system. Except for CAN, most of these early work focuses on retrieval based on one unique key such as a hash value. CAN allows for multidimensional keys in euclidean space to locate data items. All approaches, however, share the draw back that they can only retrieve items that are identified by a unique index.


The second generation of P2P systems (e.g. Mercury \cite{bharambe2004mercury}, Squid \cite{schmidt2008squid}, and Znet\cite{shu:2005}) introduced the support for more complex, multidimensional, and range queries. This enabled searches like \textit{Find persons age $\geq$ 10 and age $\leq$ 20 and gender = female}. These systems enable search in multidimensional space, where Data locality is usually achieved by dimension reduction techniques, such as space filling curves (e.g. \cite{ganesan2004one}). A general problem is that range queries might be too restricted in cases with sparse data. For example, if there are very few results for the above mentioned query, results for persons slightly older than 20 years would also be interesting for the user.



This gap was filled by research centered around nearest neighbor queries for P2P systems, like pSearch \cite{tang2003psearch} and Semantic Small World \cite{Li:2004}. The main idea is to provide nearest neighborhood search for multidimensional queries. Most work focuses on selecting important dimensions \cite{muller2003,li:2008} or methods to form an overlay network that connects nodes with similar information \cite{witschel:2005}. Just like in this paper, some of these approaches also form a small world topology \cite{Li:2004} that has a small network diameter, which makes each node reachable with only a few hops and enables efficient routing. There also exists work that relies on a predefined similarity metric, e.g. the euclidean distance, and retrieves the k nearest data items in a large collection of high dimensional data \cite{Malkov:2014,raey,batko2005scalable,li:2008}.




All these approaches have been designed to retrieve items that are explicitly defined by matching a specific identifier (id, hash value), fall in a specific range of a set of attributes, or are close to a given query. In order to retrieve knowledge this notion has to be extended by some sort of confidence metric that can take the quality of available knowledge (information) into account. 
Such a confidence metric needs to express the expertise of a node, reflecting for example that it holds a lot of similar information \cite{bach:2015} or can do reliable prediction. In our previous work \cite{bach:2015} we have tackled this issue for knowledge modeled as N-Dimensional point-clouds. We proposed a point-cluster-based confidence metric that took the variance and number of points in each cluster as an indicator of quality into account. 


To the best of our knowledge, there is no peer to peer based approach that is specifically designed so search for knowledge in graphical machine learning models.

\section{Conclusion and Future Work}

In this paper we have stressed the importance of machine learning at the edge of the network. We argued that with an increasing amount of fog computing devices carrying specialized machine learning hardware knowledge becomes inherently distributed. In this setting we defined and tackled the problem of finding and retrieving specific knowledge. We showed that our  aggregation based routing approach can retrieve specific knowledge with over 95\% accuracy even if it was learned in many different contexts. 

We think that the field of distributed knowledge management is in its infancy and will rapidly gain importance.
With our generic notion of predicting variables and context variables our retrieval strategy is flexible and can be adapted to many future application scenarios in health care, manufacturing, and urban mobility.

\section*{Acknowledgment}
The authors would like to thank the European Union's Seventh Framework Programme for partially funding this research through the ALLOW Ensembles project (project 600792).


\end{document}